\documentclass[11pt]{article}  
\usepackage{graphicx,floatflt,amssymb}    
\usepackage{epsfig,axodraw}    
\usepackage{graphics}    
\usepackage{psfrag}    
\newcommand{\ba}{\begin{array}}     
\newcommand{\ea}{\end{array}}     
\newcommand{\bd}{\begin{displaymath}}     
\newcommand{\ed}{\end{displaymath}}     
\newcommand{\be}{\begin{equation}}     
\newcommand{\ee}{\end{equation}}     
\newcommand{\bea}{\begin{eqnarray}}     
\newcommand{\eea}{\end{eqnarray}}

\newcommand{\rpv}{\mbox{$\not \hspace{-0.10cm} R_p$ }}

\def\beq{\begin{equation}}

\def\eeq{\end{equation}}

\def\bea{\begin{eqnarray}}

\def\eea{\end{eqnarray}}

\def\bq{\begin{quote}}

\def\eq{\end{quote}}

\def\gappeq{\mathrel{\rlap {\raise.5ex\hbox{$>$}}

{\lower.5ex\hbox{$\sim$}}}}

\def\lappeq{\mathrel{\rlap{\raise.5ex\hbox{$<$}}

{\lower.5ex\hbox{$\sim$}}}}

\def\bea{\begin{eqnarray}}   

\def\eea{\end{eqnarray}}

\def\Journal#1#2#3#4{{#1} {\bf #2}, #3 (#4)} 
 
\def\NPB{{\em Nucl. Phys.} B} 
\def\PLB{{\em Phys. Lett.}  B} 
\def\PRL{\em Phys. Rev. Lett.} 
\def\PRD{{\em Phys. Rev.} D} 
 
\def\PR{\em Phys. Rep.}  

\def\JHEP{\em JHEP}  

\begin{document}     
\vspace*{-0.5in}     
\renewcommand{\thefootnote}{\fnsymbol{footnote}}     
\begin{flushright}     
LPT Orsay/06-27 \\     
\end{flushright}     
\vskip 5pt

\begin{center}     
{\Large {\bf An origin for small neutrino masses in the NMSSM}}
\vskip 25pt     
{\bf Asmaa Abada $^{1,}$\footnote{E-mail
 address: abada@th.u-psud.fr}}, 
{\bf Gr\'egory Moreau $^{2,}$\footnote{E-mail
 address: greg@cftp.ist.utl.pt}}
\vskip 10pt
$^1${\it Laboratoire de Physique Th\'eorique, Universit\'e
de Paris-sud XI} \\ 
{\it B\^atiment 210, 91405 Orsay, France} \\ 
$^2${\it CFTP, Departamento de F\'{\i}sica,
Instituto Superior T\'ecnico} \\ 
{\it Av. Rovisco Pais 1,
1049-001 Lisboa, Portugal} \\ 

\normalsize     
\end{center}

\begin{abstract}  
 
We consider the Next to Minimal Supersymmetric Standard Model (NMSSM) which provides
a natural solution to the so-called $\mu$ problem by introducing a new gauge-singlet 
superfield $S$.  
We realize that a new mechanism 
of neutrino mass suppression, based on the R-parity violating bilinear terms $\mu_i L_i H_u$ 
mixing neutrinos and higgsinos, arises within the NMSSM, 
offering thus an original solution to the neutrino mass problem (connected to the 
solution for the $\mu$ problem). We 
generate realistic (Majorana) neutrino mass values without requiring any strong hierarchy amongst 
the fundamental parameters, in contrast with the alternative models. In particular, the 
ratio $| \mu_i / \mu |$ can reach $\sim 10^{-1}$, unlike in the MSSM where it has to be 
much smaller than unity. We check that the obtained parameters also satisfy the collider 
constraints and internal consistencies of the NMSSM. The price to pay for this new 
cancellation-type mechanism of neutrino mass reduction is a certain fine tuning, which 
get significantly improved in some regions of parameter space. Besides, we discuss the 
feasibility of our scenario when the R-parity violating bilinear terms have a common origin 
with the $\mu$ term, namely when those are generated via a VEV of the $S$ scalar component 
from the couplings $\lambda_i S L_i H_u$. Finally, we make comments on some specific 
phenomenology of the NMSSM in the presence of R-parity violating bilinear terms.
\\
\vskip 5pt \noindent  
\texttt{PACS Ns: 11.30.Fs, 12.60.Jv, 14.60.Pq, 14.80.Ly} \\  
\texttt{Keywords: NMSSM, Neutrino Physics, R-parity violation}  

\end{abstract}

\newpage

\renewcommand{\thesection}{\Roman{section}}  
\setcounter{footnote}{0}  
\renewcommand{\thefootnote}{\arabic{footnote}}

\section{Introduction}
\label{intro}

The most severe theoretical drawback of the Standard Model (SM) is probably 
the gauge hierarchy problem (see for example \cite{Susskind}). In well defined supersymmetric extensions 
of the SM, the property of cancellation of quadratic divergences allows to address this problem. With regard
to the field content, the most economical candidate for such a realistic extension is the Minimal Supersymmetric
Standard Model (MSSM). Nevertheless, within the MSSM, there are two unexplained hierarchies. 
\\ The first one is intrinsic to supersymmetric models: it is named as the $\mu$ problem \cite{mu}. It arises 
from the presence of a mass ($\mu$) term for the Higgs fields in the superpotential. The only two natural values 
for this $\mu$ parameter are either zero or the Planck energy scale. While the former value is excluded by experiments
as it gives rise to the unacceptable existence of an axion, the latter one reintroduces the gauge hierarchy problem.
\\ The other hierarchy with an unknown origin is the one existing between the small neutrino masses and 
the electroweak symmetry breaking scale ($\sim 100 \mbox{GeV}$). Indeed, during last years, neutrino oscillation
experiments have confirmed that neutrinos are massive. Furthermore, the additional results, extracted from 
tritium beta decay experiments and cosmological data, indicate that the values of absolute neutrino 
masses are typically smaller than the $\mbox{eV}$ scale.

In this paper, 
we propose a supersymmetric scenario which has the virtue of addressing simultaneously 
both of these hierarchy questions: the $\mu$ value naturalness {\it and} the neutrino mass 
smallness. A nice feature of our scenario is that the mechanisms explaining the two hierarchy origins 
are connected, since they involve the same additional gauge-singlet superfield, providing thus
a common source to the solutions of these two independent problems. 
\\ Our framework is the Next to Minimal Supersymmetric Standard Model (NMSSM) \cite{NMSSM}
\footnote{The phenomenology of the NMSSM was studied, for instance, in \cite{NMpheno}.}. 
The NMSSM provides an 
elegant solution to the $\mu$ problem through the introduction of a new gauge-singlet superfield $S$ 
entering the scale invariant superpotential. The scalar component of $S$ acquires naturally a Vacuum 
Expectation Value (VEV) of the order of the supersymmetry breaking scale, generating an effective $\mu$ 
parameter of order of the electroweak scale. Another appealing feature of the NMSSM is to soften the
``little fine tuning problem'' of the MSSM \cite{littleFT}. The introduction of suitable
non-renormalizable operators \cite{operators} can avoid the possibility of a cosmological domain wall 
problem \cite{wall}. There exist different explanations for a $\mu$ value of order of the electroweak
scale, but those arise in extended frameworks.
\\ In supersymmetric extensions of the SM, there exist coupling terms violating the so-called R-parity 
symmetry \cite{rpar1,rpar2}  
which acts on fields like $(-1)^{3B+L+2S}$, $B$, $L$ and $S$ being respectively
the Baryon number, Lepton number and Spin. From a purely theoretical point of view, these terms must 
be considered, even if some phenomenological limits apply on the R-parity violating (\rpv)
coupling constants \cite{reviewsD,reviewsB,HalfWay,Marc,PhysRep}. 
As a matter of fact, these terms are supersymmetric, gauge invariant
and some of them are renormalizable.  
Moreover, from the points of view of scenarios with discrete gauge symmetries \cite{Ibanez}, 
Grand Unified Theories (GUT) \cite{AlignI}-\cite{GUT1} as well as string theories \cite{string}, 
there exists no fundamental argument against the violation of the R-parity symmetry \cite{reviewsD}. 
In the present work, we consider the `bilinear' R-parity violating term
$H_u L$ appearing in the superpotential, $H_u$ and $L$ being respectively the up Higgs and
lepton doublet superfields. The existence/influence of the other \rpv terms will also be discussed. 
This bilinear interaction has been recently considered within the NMSSM context \cite{Chemtob}.
In particular, this type of interaction, which breaks the lepton number, mixes the higgsino and 
neutrinos together so that the neutrino field picks up a Majorana mass \cite{MixNeut} (the generation 
of such a neutrino mass requires two units of $L$ violation). Hence, no
additional right handed neutrino has to be introduced in order to generate a non-vanishing neutrino 
mass term.
\\ In our scenario, the smallness of absolute neutrino mass scale, with respect to electroweak scale, 
finds an origin in the following sense: the neutrino field acquires a mass of the $\mbox{eV}$ order  
without requiring any high hierarchy, 
like the usual strong hierarchy among the fundamental parameters, namely between the NMSSM 
parameters ($\lambda$ or $\mu$, as we will see later) and the \rpv coupling constants 
($\lambda_i$ or $\mu_i$, respectively). The price that
one must pay here in order to suppress the neutrino mass is a certain fine tuning on some NMSSM
parameters. However, this fine tuning can be greatly softened in specific regions of parameter 
space, since several NMSSM parameters enter the effective neutrino mass expression and some of them 
through power-law dependence.  
  
There exist mainly two supersymmetric alternatives to our scenario: two 
other kinds of model \cite{AltI,AltII} have been suggested in order to address simultaneously the 
hierarchy questions of the $\mu$ naturalness {\it and} the neutrino lightness. We will compare the 
characteristics and numerical aspects of these two models with those of our scenario.

In next section, we study the simplest version of our scenario, taking into account 
the constraints on NMSSM parameter space issued from collider physics. For that purpose, 
we use the NMHDECAY program \cite{nmhdecay}. In Section \ref{versionII},
we discuss a version where the bilinear \rpv terms are generated via the spontaneous breaking
of a symmetry. Finally, we conclude in Section \ref{conclu}.

\section{Scenario I}
\label{versionI}

\subsection{Neutralino masses}

{\bf $\bullet$ Superpotential:}
The superpotential of the NMSSM contains two characteristic terms in addition of the Yukawa 
couplings:
\begin{equation}
W_{NMSSM}= Y^u_{ij} Q_i H_u U_j^c + Y^d_{ij} Q_i H_d D_j^c + Y^\ell_{ij} L_i H_d E_j^c 
+ \lambda S H_u H_d + \frac{1}{3} \kappa S^3 ,
\label{WNMSSM}
\end{equation}
$Y^{u,d,\ell}_{ij}$ being the Yukawa coupling constants ($i,j,k$ are flavor indexes), 
$\lambda$ and $\kappa$ dimensionless coupling constants and $Q_i$, $L_i$, $U^c_i$,
$D^c_i$, $E^c_i$, $H_u$, $H_d$, $S$ respectively the superfields for the quark 
doublets, lepton doublets, up-type anti-quarks, down-type anti-quarks, anti-leptons, 
up Higgs, down Higgs, extra singlet under the SM gauge group $SU(3)_c \times SU(2)_L 
\times U(1)_Y$. The $SU(2)_L$ product of the two Higgs doublets $H_d^T=(H_d^0,H_d^-)$ 
and $H_u^T=(H_u^+,H_u^0)$ is defined as,
\begin{equation}
H_u H_d=H_u^+ H_d^- - H_u^0 H_d^0.
\label{SU2prod}
\end{equation}
The absence of terms $H_u H_d$ as well as $S^2$ and tadpoles is insured by a suitable 
discrete symmetry. An effective $\mu$ term, $\lambda \langle s \rangle  H_u H_d$, is generated  
via a VEV for the scalar component $s$ of the singlet superfield $S$.

In addition to the above NMSSM superpotential, we first consider the bilinear \rpv 
interactions:
\begin{equation}
W_I = W_{NMSSM} + \mu_i L_i H_u ,
\label{WmI}
\end{equation}
where $\mu_i$ are dimension-one \rpv parameters. The presence of the other renormalizable
\rpv interactions, namely the trilinear \rpv interactions such as $\lambda_{i,j,k} L_i L_j E_k^c$, 
depends on the symmetries of the superpotential that one assumes. It is desirable that the 
superpotential symmetries forbid the trilinear \rpv interactions violating either the lepton 
or baryon number, or both (as does the R-parity symmetry for example), in order to guarantee 
the proton stability \cite{DreinerRoss}. In other words, the whole superpotential symmetry should 
be either a Generalized Lepton (GLP), Baryon (GBP) or Matter (GMP) Parity. In case where some 
trilinear \rpv interactions are effectively present in the theory with significant coupling constant
values, those can possibly induce direct contributions to neutrino masses through one-loop level 
diagrams involving squarks or sleptons \cite{AlignI,LoopB}
\footnote{The trilinear \rpv couplings can also induce effective masses for neutrinos propagating
in matter, via tree level squark or slepton exchanges, but the SNO results forbid these contributions 
to be dominant \cite{MoDrei}.}.

{\bf $\bullet$ Soft terms:}
In the soft supersymmetry-breaking part of the Lagrangian, 
there exist also \rpv terms \cite{Chemtob}. In the 
presence of bilinear \rpv soft terms, the electroweak 
symmetry breaking can lead to a non-vanishing VEV for 
sneutrinos, denoted $\langle \tilde \nu_i  \rangle $ and 
corresponding to a possible spontaneous breaking of R-parity. These 
VEV produce new mixings between the neutrinos and neutralinos, 
contributing then to Majorana neutrino masses. 
\\ However, the $H_d$ and $L_i$ superfields, having identical quantum numbers, can be redefined by an 
$SU(4)$ rotation on $(H_d,L_i)^T$. Under this transformation, the \rpv parameters are modified. It is 
always possible to find a basis in which either $\langle \tilde \nu_i  \rangle =0$ or $\mu_i=0$. 
Nevertheless, 
generally $\langle \tilde \nu_i  \rangle $ and $\mu_i$ do not vanish simultaneously 
\cite{PhysRep}. In the present
framework, we will consider a basis where $\langle \tilde \nu_i  \rangle =0$ and $\mu_i \neq 0$.

{\bf $\bullet$ Mass matrix:}
Therefore, within our framework, the neutralino mass terms read as,
\bea
{\cal L}^m_{\tilde \chi^0}=-\frac{1}{2} \Psi^{0^T} {\cal M}_{\tilde \chi^0} \Psi^0 + H.c. 
\label{LAGmass}
\eea
in the basis defined by $\Psi^{0^T}$ $\equiv$ 
$(\tilde B^0, \tilde W^0_3, \tilde h^0_d, \tilde h^0_u, \tilde s, \nu_i)^T$,
where $\tilde h^0_{u,d}$ ($\tilde s$) is the fermionic component of the superfield $H_{u,d}^0$ ($S$)
and the $\nu_i$ denote the neutrinos. 
In Eq.(\ref{LAGmass}), the neutralino mass matrix is given by, 
\bea
{\cal M}_{\tilde \chi^0} = 
\left( 
\begin{array}{cc}
{\cal M}_{NMSSM}  & \xi_{\rpv}^{T} \\
\xi_{\rpv} & {\bf 0}_{3 \times 3}
\end{array}
\right)
\label{CHImass}
\eea
where ${\cal M}_{NMSSM}$ is the neutralino mass matrix which 
holds in the NMSSM with conserved R-parity ($s$, $c$ standing for $\sin$, $\cos$): 
\bea
{\cal M}_{NMSSM}
=
 \left(
 \begin{array}{cccccc}
M_{1}&0 & -M_Z \ s\theta_W \ c\beta & M_Z \ s\theta_W \ s\beta & 0\\ 
0 & M_{2} & M_Z \ c\theta_W \ c\beta & -M_Z \ c\theta_W \ s\beta  &0\\
 -M_Z \ s\theta_W \ c\beta & M_Z \ c\theta_W \ c\beta & 0 &-\mu& -\lambda v_u\\
M_Z \ s\theta_W \ s\beta &-M_Z \ c\theta_W \ s\beta &-\mu&0& -\lambda v_d \\
0&0&-\lambda v_u &-\lambda v_d & 2 \kappa \langle s \rangle  \\
  \end{array}
 \right)\ ,
\label{NMSSMmass}
\eea
$\xi_{\rpv}$ is the \rpv part of the matrix mixing neutrinos and neutralinos,
\bea
\xi_{\rpv} = 
\left( 
\begin{array}{ccccc}
0 & 0 & 0 & \mu_1 & 0  \\
0 & 0 & 0 & \mu_2 & 0  \\
0 & 0 & 0 & \mu_3 & 0  
\end{array}
\right)
\label{RPVmass}
\eea
$M_1$ ($M_2$) is the soft supersymmetry breaking mass of the bino (wino), $M_Z$ the $Z^0$ boson 
mass, $\theta_W$ the electroweak angle, $\tan \beta=v_u/v_d=\langle h^0_u \rangle /\langle h^0_d \rangle $ ($h^0_{u,d}$ being the scalar 
component of $H_{u,d}^0$) and 
\begin{equation}
\mu = \lambda \langle s \rangle .
\label{mu}
\end{equation}

{\bf $\bullet$ Parameters:}
In this scenario, the independent parameters in the neutralino sector can be chosen
as being the following set of variables, 
\begin{equation}
\lambda, \ \kappa, \ \tan \beta, \ \mu, \ M_1, \ M_2.
\label{parameters}
\end{equation}
We take these variables as free parameters at the electroweak scale. We adopt the convention of signs
in which $\lambda > 0$, $\tan \beta > 0$ (without loss of generality) whereas $\kappa$ and $\mu$ can 
take positive or negative values. Finally, we assume that $\lambda$, $\kappa$ and the soft supersymmetry 
breaking parameters are real.

\subsection{Effective neutrino mass}
\label{effnumass}

{\bf $\bullet$ Mass expression:}
We restrict ourselves to the case $|\mu_i/\mu|<10^{-1}$ and to some parameter values (in particular 
sufficiently large $M_{1,2}$) such that the neutrino-neutralino mixing terms remain much smaller 
than the neutralino masses. Hence, the effective neutrino mass matrix is given in a good 
approximation by the following formula, having a ``see-saw'' type structure,
\begin{equation}
m_{\nu} = - \xi_{\rpv} \ {\cal M}_{NMSSM}^{-1} \ \xi_{\rpv}^T.
\label{seesaw}
\end{equation}
We have checked, through a comparison with an exact numerical diagonalization, that this block 
form expression represents systematically a good approximation for all the points of parameter 
space that we consider in this work. From Eq.(\ref{NMSSMmass}), Eq.(\ref{RPVmass}) and 
Eq.(\ref{seesaw}), we deduce an analytic expression for the effective Majorana neutrino mass 
matrix:
\begin{equation}
m_{\nu i j} = \mu_i \mu_j \frac{M_1 M_2 (2 \kappa \mu / \lambda)}{Det({\cal M}_{NMSSM})}
\bigg (
\frac{(\lambda v_u)^2}{2 \kappa \mu / \lambda} + M_Z^2
[ \frac{\sin^2\theta_W}{M_1} + \frac{\cos^2\theta_W}{M_2} ] \cos^2 \beta
\bigg ),
\label{MnuEFF}
\end{equation}
$Det({\cal M}_{NMSSM})$ being the determinant of the matrix (\ref{NMSSMmass}).

{\bf $\bullet$ Origin of smallness:}
We observe on neutrino mass matrix (\ref{MnuEFF}) that the overall factor can be significantly
suppressed if the two terms in brackets compensate each other by taking opposite signs and
approximately equal absolute values. This means that neutrino mass eigenvalues can be affected 
by an important suppression factor. This neutrino mass suppression has a different and new origin 
with respect to the other possible suppression coming from the smallness of ratio $|\mu_i/\mu|$ 
({\it c.f.} Eq.(\ref{MnuEFF})). The smallness of ratio $|\mu_i/\mu|$ cannot constitute a physical 
interpretation to the smallness of neutrino mass scale compared to electroweak scale, 
in the sense that an other (unexplained) mass hierarchy is introduced.

\begin{figure}[t]\unitlength1mm
\SetScale{2.8}
\begin{boldmath}
\begin{tabular}{c c c}
\begin{picture}(80,80)(0,-10)
\Line(0,0)(15,0)\Line(65,0)(15,0)
\Line(80,0)(65,0)
\DashLine(25,0)(25,30){2}
\Text(22,-4.5)[c]{$\widetilde  {h}^0_{u,d}$}
\DashLine(55,0)(55,30){2}
\Text(60,-4.5)[c]{$\widetilde  {h}^0_{u,d}$}
\Text(52.4,29.6)[l]{$\times$}
\Text(55,35)[c]{$\langle {h}^0_{u,d} \rangle $}
\Text(23,29.6)[l]{$\times$}
\Text(25,35)[c]{$\langle {h}^0_{u,d} \rangle $}
\Text(73,0)[r]{$<$}
\Text(61,0)[r]{$>$}
\Text(-2,0)[r]{$\nu_i$}
\Text(8,0)[r]{$>$}
\Text(22,0)[r]{$<$}
\Text(80,0)[l]{$\nu_j$}
\Text(40,-6)[c]{$\widetilde{B^0},\widetilde{W}^0_3$}
\Text(40,-12)[c]{(a)}
\Text(40,5)[c]{$M_{1,2}$}
\Text(33,0)[r]{$>$}
\Text(50,0)[r]{$<$}
\Text(40,0)[c]{$\times$}
\Text(15,0)[c]{x}
\Text(65,0)[c]{x}
\Text(15,5)[c]{$\mu_i$}
\Text(65,5)[c]{$\mu_j$}
\end{picture}
&\strut\hspace*{2mm}&
\begin{picture}(80,80)(0,-10)
\Line(0,0)(15,0)\Line(65,0)(15,0)
\Line(78,0)(65,0)
\DashLine(25,0)(25,30){2}
\Text(22,-4.5)[c]{$\widetilde  {h}^0_{u,d}$}
\DashLine(55,0)(55,30){2}
\Text(60,-4.5)[c]{$\widetilde  {h}^0_{u,d}$}
\Text(52.6,29.6)[l]{$\times$}
\Text(55,35)[c]{$\langle {h}^0_{u,d} \rangle $}
\Text(23.2,29.6)[l]{$\times$}
\Text(25,35)[c]{$\langle {h}^0_{u,d} \rangle $}
\Text(73,0)[r]{$<$}
\Text(61,0)[r]{$>$}
\Text(-1,0)[r]{$\nu_i$}
\Text(8,0)[r]{$>$}
\Text(22,0)[r]{$<$}
\Text(80,0)[l]{$\nu_j$}
\Text(40,-6)[c]{$\widetilde{s}$}
\Text(33,0)[r]{$>$}
\Text(50,0)[r]{$<$}
\Text(40,5)[c]{$2\kappa \mu/\lambda$}
\Text(40,0)[c]{$\times$}
\Text(15,0)[c]{x}
\Text(65,0)[c]{x}
\Text(15,5)[c]{$\mu_i$}
\Text(62,5)[c]{$\mu_j$}
\Text(40,-12)[c]{(b)}
\end{picture}
\end{tabular}
\end{boldmath}
\caption{(a) Feynman diagram for the typical contribution to the Majorana neutrino masses arising 
in the MSSM from mixing with neutralinos (see text for notations of fields and parameters).
The effective mass affecting the two vertex is of type $m=\pm M_Z t(\theta_W) t(\beta)$, 
where $t(x)$ is equal to either $\sin x$ or $\cos x$. 
A cross indicates either a mass insertion or 
a VEV. The arrows show the flow of momentum for associated propagators.
(b) Feynman diagram for the additional type of contribution to the Majorana neutrino masses 
arising in the NMSSM from mixing with neutralinos. The mass parameter at the two vertex 
is there $m= - \lambda v_{u,d}$.}
\protect\label{fig:diagramMSSM}
\end{figure}
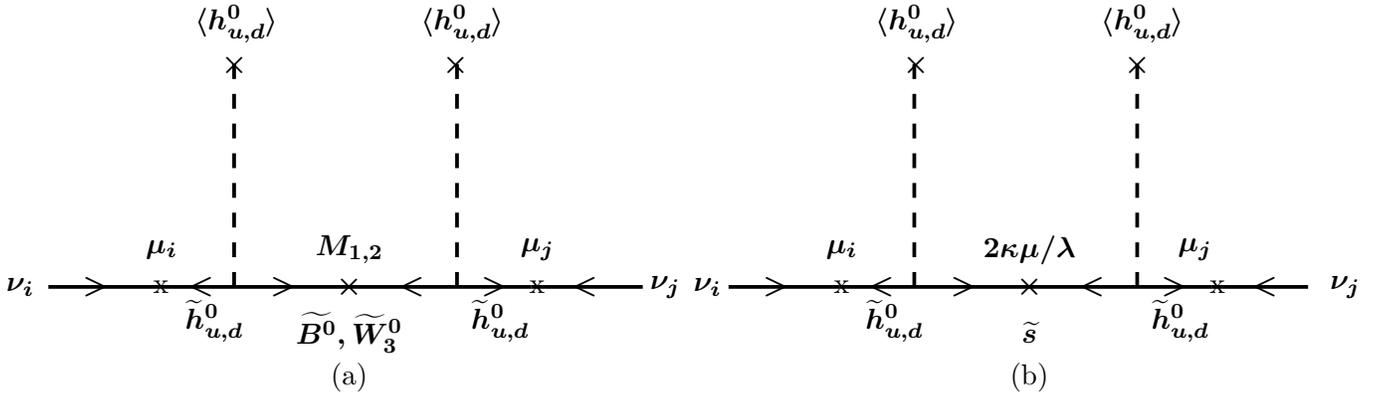

Let us understand this possible cancellation in Eq.(\ref{MnuEFF}) 
from a diagramatic point of view. In Fig.(\ref{fig:diagramMSSM}), 
we present the two characteristic diagrams of neutralino mass 
contributions to Majorana neutrino mass terms. We see that the first term in brackets of neutrino 
mass expression (\ref{MnuEFF}) corresponds to the exchange of a gaugino shown in Fig.(\ref{fig:diagramMSSM})(a). 
Indeed, this first term is of the type $\mu_i \mu_j (m^2 / M)$, where $m$ is the typical mass entering 
at the two vertex linked to a Higgs VEV and $M=M_{1,2}$ is the gaugino   
mass. The second term in brackets of formula (\ref{MnuEFF}) is associated to the exchange of the singlino 
$\tilde s$ shown in Fig.(\ref{fig:diagramMSSM})(b): this term is also of the type $\mu_i \mu_j (m^2 / M)$, 
where $m$ is the typical mass at vertex with a Higgs VEV and now 
$M= 2 \kappa \langle s \rangle  = 2 \kappa \mu / \lambda$ is the singlino mass 
(see Eq.(\ref{NMSSMmass}) and Eq.(\ref{mu})).
In conclusion, an approximate cancellation between the two terms in the brackets entering neutrino mass 
expression (\ref{MnuEFF}) would represent a compensation between the exchanges of a gaugino and a singlino.

The price of this neutrino mass suppression 
is a certain amount of fine tuning on some NMSSM parameter values. The $\lambda$ parameter faces the most 
important fine tuning. Nevertheless, this fine tuning is significantly reduced when $| \mu |$, 
$M_{1,2}$ and $\tan \beta$ increase. In next section, we discuss this aspect more precisely 
and quantitatively.

\subsection{Numerical results}
\label{NumResI}

{\bf $\bullet$ Flavors:}
In the discussion of the main features of our scenario, we will concentrate on the 
case of one neutrino flavor, for simplification reasons. The treatment of the realistic 
three-flavor case requires the calculation of loop contributions to the neutrino mass
matrix, via the Grossman-Haber diagrams \cite{GroHab} which kill the 
degeneracy in neutrino mass spectrum. 
Indeed, at the tree level, only one of the three neutrino eigenstates 
obtains a non-vanishing mass eigenvalue (from the mixing with neutralinos), a scheme which 
conflicts with present data as we know that solar and atmospheric neutrino data
require at least two non-zero eigenvalues \cite{Valle}. The possibility, that the combined tree and 
one-loop contributions could account for the observed data on three neutrino masses and three
leptonic mixing angles, has been investigated extensively in the context of the MSSM 
\cite{MSSMrpvA}-\cite{MSSMrpvB}.
Such a three-flavor global fit of all neutrino data at the loop level within the NMSSM
is beyond the scope of our study. Nevertheless, we comment that, as in the MSSM, the soft supersymmetry
breaking interactions (like $B_i h^0_u \tilde \nu_i$) \cite{GroHab}, the cancellations between 
contributions involving the Higgs sector modes \cite{cancel} as well as the sneutrino mass splittings 
should play a crucial r\^ole in the computations for loop amplitudes of neutrino masses.

At one flavor, we see from Eq.(\ref{MnuEFF}) that the neutrino mass can be written as, 
\begin{equation}
m_{\nu} = \frac{\mu_1^2}{M_{SUSY}},
\label{MnuOne}
\end{equation}
where $M_{SUSY}$ is an effective mass depending on the supersymmetry (breaking) parameters.
In the three-flavor case, the only non-vanishing neutrino mass eigenvalue at tree level reads as,
\begin{equation}
m^{high}_{\nu} = \frac{\mu_1^2+\mu_2^2+\mu_3^2}{M_{SUSY}}.
\label{MnuThree}
\end{equation}
At loop level, the two other neutrino mass eigenvalues receive loop contributions. 
We consider that the largest neutrino mass eigenvalue remains given by $m^{high}_{\nu}$ 
in Eq.(\ref{MnuThree}) to a good approximation. By consequence, the neutrino mass $m_{\nu}$ in 
Eq.(\ref{MnuOne}) that we will consider at one flavor, is approximately equal to the largest 
neutrino mass eigenvalue at three flavors, namely $m^{high}_{\nu}$ in Eq.(\ref{MnuThree}), 
for $\mu_{2,3}^2 \lesssim \mu_1^2$.

{\bf $\bullet$ Neutrino mass constraints:}
Let us summarize the existing experimental constraints on the largest neutrino mass eigenvalue
$m^{high}_{\nu}$.
First, a three-flavor global fit analysis, including the results from solar, atmospheric, reactor (KamLAND 
and CHOOZ) and accelerator (K2K) experiments, leads to the following intervals at the $4 \sigma$ 
level \cite{Valle}:
$6.8 \leq \Delta m_{21}^2 \leq 9.3 \ \ \ [10^{-5} \mbox{eV}^2]$ and
$1.1 \leq \Delta m_{31}^2 \leq 3.7 \ \ \ [10^{-3} \mbox{eV}^2]$,  
$\Delta m_{21}^2 \equiv m_{\nu_2}^2-m_{\nu_1}^2$ and $\Delta m_{31}^2 \equiv m_{\nu_3}^2-m_{\nu_1}^2$ 
being the differences of squared neutrino mass eigenvalues. Hence, the largest neutrino mass eigenvalue
is larger than about $\sqrt{3 \ 10^{-3} \mbox{eV}^2}$, which can be written as,
\begin{equation}
0.05 \mbox{eV} \lesssim m^{high}_{\nu}.
\label{bdINF}
\end{equation} 
We now turn to the current upper experimental limits on absolute neutrino mass scales. We first
consider the limits extracted from the tritium beta decay experiments \cite{107,108,109} which
are independent of the nature of neutrino mass (Majorana or Dirac). The data provided by the 
Mainz \cite{108} and Troitsk \cite{109} experiments give rise to the bounds (at $95 \% \ C.L.$):
$m_\beta \leq 2.2 \ \mbox{eV} \ \ \ \mbox{[Mainz]}$ and
$m_\beta \leq 2.5 \ \mbox{eV} \ \ \ \mbox{[Troitsk]}$,
where the effective mass $m_\beta$ is defined by $m^2_\beta = \sum_{i=1}^3 |U_{ei}|^2 m_{\nu_i}^2$,
$U_{ei}$ being the leptonic mixing matrix. This matrix is parameterized 
by the three mixing angles $\theta_{12}$, $\theta_{23}$ and $\theta_{13}$ 
which are constrained to lie in the ranges \cite{Valle}:
$0.21 \leq \sin^2 \theta_{12} \leq 0.41$, 
$0.30 \leq \sin^2 \theta_{23} \leq 0.72$ and
$\sin^2 \theta_{13} \leq 0.073$. 
From the above constraints, we deduce that the largest neutrino 
mass eigenvalue is bounded from above typically by
\begin{equation}
m^{high}_{\nu} \lesssim 1 \mbox{eV}.
\label{bdSUP}
\end{equation} 
Secondly, the cosmological data from WMAP and 2dFGRS galaxy survey \cite{cosmobound} place the
following bound (depending on cosmological priors):
$\sum_{i=1}^3 m_{\nu_i} \lesssim 0.7 \mbox{eV}$.
This bound gives rise to an upper limit on the largest 
neutrino mass eigenvalue which is of the same order of magnitude as in Eq.(\ref{bdSUP}).

As we have discussed above, the largest neutrino mass eigenvalue $m^{high}_{\nu}$, 
at three flavors, is approximately equal to the neutrino mass $m_{\nu}$, at one flavor.
One thus concludes from the typical bounds (\ref{bdINF}) and (\ref{bdSUP}) that
\begin{equation}
0.1 \mbox{eV} \lesssim m_{\nu} \lesssim 1 \mbox{eV}. 
\label{bdTOT}
\end{equation} 

\begin{table}[t]
\begin{center}
\begin{tabular}{c|c|c|c|c|c|c|c}
 & $\kappa$ & $\mu$ & $\tan \beta$ & $M_1$ & $M_2$ & $\mu_1$ & $\lambda$  
\\
&  & $\mbox{[GeV]}$ &  & $\mbox{[TeV]}$ & $\mbox{[TeV]}$ & $\mbox{[GeV]}$ &
\\
\hline
A & 0.05 &  -400 &  54   & 5 & 5 & $10$ & $(9.0969 - 9.105) \ 10^{-3}$ 
\\
&  &   &    &  &  & $10^{-1}$ & $(1.43 - 2.8) \ 10^{-2}$ 
\\
\hline
B & 0.05 &  -300 &  50   & 1 & 1 & $10$ & $(1.48759 - 1.48771) \ 10^{-2}$ 
\\
&  &   &   & & & $10^{-1}$ & $(1.613 - 2.31) \ 10^{-2}$ 
\\
\hline
C & 0.15 &  -300 &  30   & 5 & 5 & $10$ & $(1.76374 - 1.764) \ 10^{-2}$ 
\\
& &  &   &  &  & $10^{-1}$ & $(2.014 - 3.2) \ 10^{-2}$ 
\\
\hline
D & 0.2 &  -200 &  50   & 3 & 4 & $10$ & $(1.3321 - 1.3323) \ 10^{-2}$ 
\\
& &  &   &  &  & $10^{-1}$ & $(1.51 - 2.35) \ 10^{-2}$ 
\\
\hline
E & -0.1 &  300 &  50   & 3 & 3 & $10$ & $(1.29955 - 1.2999) \ 10^{-2}$ 
\\
&  &  &   &  &  & $10^{-1}$ & $(1.585 - 2.72) \ 10^{-2}$ 
\end{tabular}
\caption{Sets (A,\dots,E) of values, for the parameters entering the whole neutralino mass matrix 
(\ref{CHImass}), which reproduce the correct neutrino mass. The two values of parameter $\lambda$ 
correspond to the neutrino masses $m_{\nu}= 0.1 \mbox{eV} - 1 \mbox{eV}$ ($m_{\nu}$ being defined 
via Eq.(\ref{MnuEFF})), respectively. As the one-flavor case is considered here, the flavor index
$i$ of \rpv parameter $\mu_i$ takes only the value $i=1$ (as in Eq.(\ref{MnuOne})).}
\label{tab:parI}
\end{center}
\end{table}

{\bf $\bullet$ Neutrino mass suppression:}
In Table \ref{tab:parI}, we present characteristic points of parameter space for which
the neutrino mass (\ref{MnuEFF}) at one flavor, namely $m_{\nu}$ ({\it c.f.} Eq.(\ref{MnuOne})), 
is equal to $0.1 \mbox{eV}$ and $1 \mbox{eV}$, in order to cover the typical range of values 
allowed by experimental results (see Eq.(\ref{bdTOT})).

In fact, for each of the sets of parameters shown in Table \ref{tab:parI}, the $\lambda$ value
is determined as a function of the other parameters through the formula (\ref{MnuEFF}) for neutrino 
mass. In other terms, the relation (\ref{MnuEFF}) fixes one of the parameters (as the $m_{\nu}$ 
value is given) that we choose to be $\lambda$. The $\lambda$ values are written with the accuracy
necessary to obtain the wanted neutrino mass. This accuracy reflects two aspects: the fact that it is 
$\lambda$ that we determine as a function of the other parameters, and, the fine tuning needed on 
$\lambda$ (which will be discussed in more details in next table). As already said, $\lambda$
is the quantity that suffers from the most important fine tuning.

Let us discuss the physical meaning of results presented in Table \ref{tab:parI}. 
We remark that for the signs of parameters systematically chosen in this table
(note the different sign configuration for last point E), the approximate cancellation 
between the two terms in brackets entering neutrino mass expression (\ref{MnuEFF}) 
is effective as these two terms possess opposite signs. The first possibility is that this 
cancellation is only partially responsible for the neutrino mass suppression relatively to 
the electroweak scale: this is the case for all the points in this table with 
$| \mu_1/\mu | \simeq 10^{-3}$ ($\mu_1 = 10^{-1} \mbox{GeV}$). In that case, the 
suppression of neutrino mass is also due to the hierarchy introduced between the \rpv 
parameter $\mu_1$ and the effective $\mu$ quantity. The other possibility is that the above
cancellation constitutes the main mechanism suppressing the neutrino mass: this is the case 
for the points with $| \mu_1/\mu | \simeq 10^{-1}$ ($\mu_1 = 10 \mbox{GeV}$). In that case, 
the necessary neutrino mass suppression is achieved without introducing any new strong 
hierarchy among the parameters of the theory. 

This result, that the smallness of neutrino mass can be mainly
due to a compensation between two contributions exchanging a gaugino and a singlino, is 
one of the major and new results of our paper.

{\bf $\bullet$ $\mu$ naturalness:}
Let us comment about the parameter values taken in Table \ref{tab:parI}. Motivated by arguments of 
naturalness, one may wish to restrict to $\langle s \rangle  \lesssim 10 \mbox{TeV}$, which translates ({\it c.f.} 
Eq.(\ref{mu})) into the condition $|\mu| [\mbox{GeV}] \times 10^{-4} \lesssim \lambda$. This condition 
is satisfied by the values obtained in Table \ref{tab:parI}. Besides, the absence of Landau singularities,
for $\lambda$, $\kappa$ and the Yukawa coupling constants $Y^b,Y^t$ below the GUT energy scale, imposes 
\cite{darkNMSSM} the typical bounds on NMSSM parameters: $\lambda \lesssim 0.75$, $|\kappa| \lesssim 0.65$ 
and $1.7 \lesssim \tan \beta \lesssim 54$. All the parameter values in Table \ref{tab:parI} satisfy 
these bounds. Finally, the various values of $\mu$ in this table have been chosen such that 
$|\mu| \gtrsim 100 \mbox{GeV}$, in order to safely respect the LEP bound on the lightest chargino 
mass: $m_{\tilde \chi_1^+}>103.5 \mbox{GeV}$ \cite{LEPchargino}.

In addition, we have checked that the parameter sets presented in Table \ref{tab:parI} belong well
to some regions of the NMSSM parameter space which are compatible with the various theoretical consistencies 
and experimental constraints. For that purpose, we have performed a scan, by using the Fortran code NMHDECAY 
\cite{nmhdecay}, in order to test the following parameter ranges: $0.009< \lambda <0.02$, $0.05< | \kappa | <0.2$,
$30< \tan \beta <54$ and $100\mbox{GeV}< | \mu | <400\mbox{GeV}$. This scan was done simultaneously with a scan 
over $-1\mbox{TeV}< A_\lambda <1\mbox{TeV}$ and $-1\mbox{TeV}< A_\kappa <1\mbox{TeV}$, where $A_\lambda$ and 
$A_\kappa$ are the trilinear soft supersymmetry breaking parameters (entering the NMSSM Lagrangian via the terms 
$\lambda A_\lambda s h_u h_d$ and $(1/3) \kappa A_\kappa s^3$) which do not affect the neutralino mass matrix 
(\ref{CHImass}). Precisely, 
the NMHDECAY program has allowed us to check that \cite{nmhdecay} {\it (i)} the physical 
minimum of the scalar potential is deeper than the local unphysical minima with $\langle h^0_{u,d} \rangle =0$ and/or $\langle s \rangle =0$
{\it (ii)} the running couplings $\lambda$, $\kappa$, $Y^b$ and $Y^t$ do not encounter a Landau singularity 
{\it (iii)} the experimental constraints from LEP in the neutralino, chargino and Higgs sectors are 
effectively satisfied.

The consistency of using the code NMHDECAY (which strictly speaking deals with the pure NMSSM) in our present 
scenario is justified by the following argument.
The presence of the additional bilinear \rpv term $\mu_i L_i H_u$ in the superpotential (see Eq.(\ref{WmI})), 
that we have supposed, does not automatically modifies the Higgs potential of the NMSSM at tree level. 
Indeed, the term $\mu_i^2 |h_u|^2$ in the Higgs potential, coming from the bilinear \rpv term, can be 
reabsorbed in a redefinition of the soft Higgs mass term $m_{h_u}^2 |h_u|^2$.

\begin{table}[t]
\begin{center}
\begin{tabular}{c|c|c|c}
& $m_{\tilde \chi^0_1}$  & $m_{\tilde \chi^0_5}$  &  ${\cal F}_\lambda$  
\\
& $\mbox{[GeV]}$ & $\mbox{[GeV]}$ &
\\
\hline
A & 399 & 5002 &   $(0.9 - 9.3) \ 10^{-4}$ 
\\
& 399 & 5002 & $(2.5 - 3.2) \ 10^{-1}$   
\\
\hline
B & 295 & 2017 &  $(0.9 - 9.0) \ 10^{-5}$ 
\\
& 294 & 1862 - 1303  & $(0.7 - 2.4) \ 10^{-1}$   
\\
\hline
C & 299 & 5103 - 5102  &  $(0.2 - 1.6) \ 10^{-4}$ 
\\
& 299 & 5002 & $(1.1 - 2.8) \ 10^{-1}$ 
\\
\hline
D & 199 & 6006 - 6005 &  $(0.1 - 1.5) \ 10^{-4}$ 
\\
& 199 & 5310 - 4002 &  $(1.0 - 2.7) \ 10^{-1}$ 
\\
\hline
E & 299 & 4617 - 4616  & $(0.3 - 2.7) \ 10^{-4}$ 
\\
& 298 & 3787 - 3003 &   $(1.5 - 3.0) \ 10^{-1}$ 
\end{tabular}
\caption{Lowest [$m_{\tilde \chi_1^0}$] and highest [$m_{\tilde \chi_5^0}$] neutralino masses 
(among the six mass eigenvalues of matrix (\ref{CHImass}), 
except the neutrino mass eigenvalue $m_{\nu}$) for the points A,\dots,E of parameter space presented in 
Table \ref{tab:parI}. Together with these masses, we also show the value of fine tuning function 
${\cal F}_\lambda$ defined in the text for the $\lambda$ parameter. The two values of ($m_{\tilde \chi_5^0}$ 
and) ${\cal F}_\lambda$ correspond respectively to the two $\lambda$ values in Table \ref{tab:parI}
(leading to $m_{\nu}= 0.1 \mbox{eV} - 1 \mbox{eV}$). For each point, the first and second lines are 
respectively associated to $\mu_1=10 \mbox{GeV}$ and $\mu_1=10^{-1} \mbox{GeV}$, as in Table 
\ref{tab:parI}.}
\label{tab:FTI}
\end{center}
\end{table}

{\bf $\bullet$ Fine tuning:}
The mechanism of neutrino mass suppression presented in Section \ref{effnumass} requires a certain
amount of fine tuning. In order to discuss quantitatively this fine tuning on the $\lambda$ parameter 
(the most important fine tuning), we introduce the following ratio,
\begin{equation}
{\cal F}_\lambda 
= \bigg | \frac{\delta ln \lambda}{\delta ln m_{\nu}} \bigg | 
= \bigg | \frac{\delta \lambda / \lambda}{\delta m_{\nu} / m_{\nu}} \bigg | ,
\label{FTratio}
\end{equation} 
where $\delta m_{\nu}$ is the variation of neutrino mass associated to the variation $\delta \lambda$ 
of fundamental parameter $\lambda$, for any other parameter fixed to a certain value. The largest values 
of this quantity ${\cal F}_\lambda$ correspond to the most soft fine tuning. By using the neutrino mass 
expression (\ref{MnuEFF}), we have calculated analytically the quantity ${\cal F}_\lambda$ as a function 
of the fundamental parameters of the neutralino mass matrix.

In Table \ref{tab:FTI}, we give the values of this function ${\cal F}_\lambda$ for the points of 
parameter space presented in Table \ref{tab:parI} which generate acceptable neutrino masses through 
our cancellation mechanism. By comparing the points A and B of Table \ref{tab:FTI}, we observe, through
the values of function ${\cal F}_\lambda$, that the fine tuning get softer as $M_{1,2}$ increases. 
Similarly, the comparison of parameter sets A and C (A and D) shows that the fine tuning is significantly 
improved for larger values of $\tan \beta$ ($| \mu |$). The point E, corresponding to different signs 
of $\kappa$ and $\mu$ than for the other points, exhibits the weak dependence of fine tuning on the 
sign configurations. Finally, we remark that the fine tuning is softer for an higher neutrino mass
(second ${\cal F}_\lambda$ values in Table \ref{tab:FTI}) as well as for a smaller $|\mu_i/\mu|$ ratio
(second line for each point). The reason is that, in these two cases, the neutrino mass suppression 
mechanism, which is based on the compensation of two mass contributions, has to be less effective
(i.e. it must suppresses less the absolute neutrino mass scale).

To finish the comments about Table \ref{tab:FTI}, we mention that, for each parameter set considered, 
the largest neutralino mass eigenvalue $m_{\tilde \chi^0_5}$ is of order of the $\mbox{TeV}$ scale 
so that the gauge hierarchy problem remains addressed through the supersymmetry.

\begin{figure}[t]
\begin{center} 
\psfrag{mu}[c][c][1]{{\large $\mu \ \mbox{[GeV]}$}} 
\psfrag{tanbeta}[c][r][1]{{\large $\tan \beta$}}
\includegraphics[width=0.5\textwidth,height=6cm]{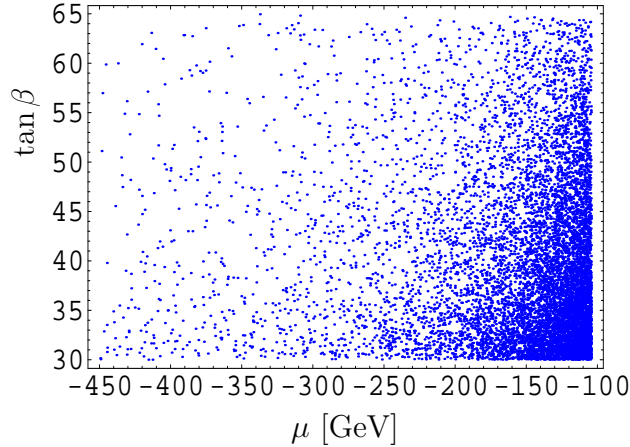}
\caption{Points in the plan $\mu$ (in $\mbox{GeV}$) versus $\tan \beta$ producing a neutrino mass
$m_{\nu}=1 \mbox{eV}$, for $\mu_1 = 1 \mbox{GeV}$, $M_1=3 \mbox{TeV}$, $M_2=4 \mbox{TeV}$ 
and values of $\lambda$ and $\kappa$ given by Fig.(\ref{fig:lakp}).} 
\protect\label{fig:mutb}
\end{center}
\end{figure}

\begin{figure}[t]
\begin{center} 
\psfrag{lambda}[c][c][1]{{\large $\lambda$}} 
\psfrag{kappa}[c][r][1]{{\large $\kappa$}}
\psfrag{pta}[c][c][1]{{\footnotesize 0.0025}}
\psfrag{ptb}[c][c][1]{{\footnotesize 0.005}}
\psfrag{ptc}[c][c][1]{{\footnotesize 0.0075}}
\psfrag{ptd}[c][c][1]{{\footnotesize 0.01}}
\psfrag{pte}[c][c][1]{{\footnotesize 0.0125}}
\psfrag{ptf}[c][c][1]{{\footnotesize 0.015}}
\psfrag{ptg}[c][c][1]{{\footnotesize 0.0175}}
\includegraphics[width=0.6\textwidth,height=6.5cm]{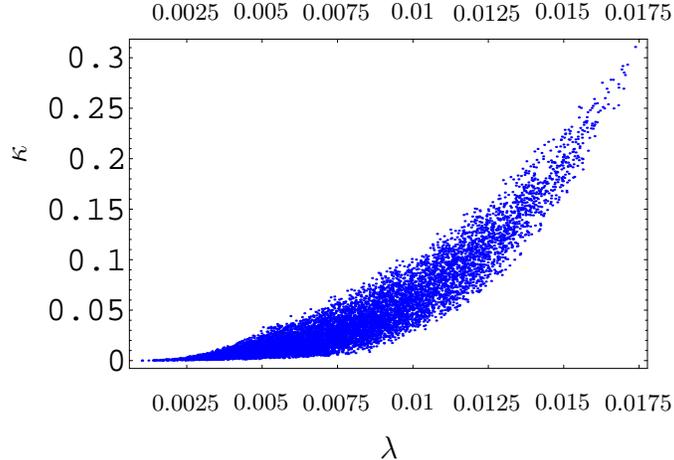}
\caption{Points in the plan $\lambda$ versus $\kappa$ producing a neutrino mass
$m_{\nu}=1 \mbox{eV}$, for $\mu_1 = 1 \mbox{GeV}$, $M_1=3 \mbox{TeV}$, $M_2=4 \mbox{TeV}$ 
and values of $\mu$ and $\tan \beta$ given by Fig.(\ref{fig:mutb}).}
\protect\label{fig:lakp}
\end{center}
\end{figure}

{\bf $\bullet$ Scans:}
In Fig.(\ref{fig:mutb}) and Fig.(\ref{fig:lakp}), we show points of the NMSSM parameter space
generating a neutrino mass at one flavor ({\it c.f.} Eq.(\ref{MnuOne})) equal to $1 \mbox{eV}$,
for $\mu_1=1\mbox{GeV}$. This $\mu_1$ value corresponds to $| \mu_1/\mu | \simeq 10^{-2}$, 
which means that, for the points presented in the two figures, the dominant effective suppression 
mechanism of neutrino mass is our cancellation mechanism (a ratio of $|\mu_i/\mu| \sim 10^{-6}$ 
is needed to obtain the entire neutrino mass reduction from the hierarchy between $\mu_i$ and $\mu$,
as we will discuss later). 

The points in Fig.(\ref{fig:mutb})-(\ref{fig:lakp}) have been obtained through a scan performed 
with the NMHDECAY program, so that they respect the experimental and theoretical constraints mentioned 
above.

These two figures show that an acceptable neutrino mass can be generated, via the considered
cancellation model, in large regions of the NMSSM parameter space. Besides, Fig.(\ref{fig:lakp}) 
exhibits a correlation between the coupling constants $\lambda$ and $\kappa$ which is characteristic
of the cancellation mechanism.

{\bf $\bullet$ Lepton flavor violation:}
In the present framework, we have shown that the experimental values of neutrino masses allow the
ratio $| \mu_i/\mu |$ to be as large as $\sim 10^{-1}$. Such a possible enhancement of $| \mu_i/\mu |$
tends to increase the amplitudes of low energy lepton flavor violating processes like $\mu \to eee$
or $\mu \to e \gamma$. Indeed, these decay processes receive tree level contributions through the mixings
of type $\mu_i \tilde h^+_u \ell_i$. We obtain, via simple estimations (as done for instance in 
\cite{adl}), 
that the experimental upper limits \cite{PDG} on the branching ratios of these decay processes are 
respected for $| \mu_i/\mu |$ values up to $\sim 10^{-2}$.

\subsection{Comparison with the other models}

Finally, we compare our scenario with existing alternative supersymmetric models. First, it has been suggested
recently \cite{AltII} that a gauge-singlet right handed neutrino $N_i^c$, added to the MSSM superfield content in 
order to generate Dirac neutrino masses (via $Y^\nu_{ij} L_i H_u N_j^c$), can also play the r\^ole of the NMSSM 
singlet S. Indeed, the scalar components of $N_i^c$ (sneutrinos) can produce an effective $\mu$ term (via 
$\lambda^i N_i^c H_u H_d$) by acquiring a VEV. In this so-called new MSSM, 
the R-parity is broken explicitly via the cubic term  
for $N_i^c$ ($(1/3)\kappa^{ijk}N_i^cN_j^cN_k^c$). The two other model-building differences of this new MSSM 
with our scenario are that, here, the added gauge-singlet $N_i^c$ comes with a flavor index and has a right 
handed chirality (in contrast with $S$). Hence, in the new MSSM, there are three distinct origins to the 
neutrino mass: the mixing with gauginos/higgsinos, the Majorana neutrino mass proportional to $\kappa^{ijk}$ and 
the Dirac neutrino mass involving the Yukawa coupling constants $Y^\nu_{ij}$. These Yukawa coupling constants 
must be of order $10^{-6}$ in order to obtain reasonable neutrino mass eigenvalues around $10^{-2}\mbox{eV}$ 
\cite{AltII}. This means that a hierarchy of $\sim 10^{-6}$ has to be introduced between the Yukawa couplings 
of the neutrinos and the top quark ($Y^t \sim 1$). This has to be contrasted with our mechanism which can produce 
acceptable neutrino masses with only a little hierarchy of $|\mu_i/\mu| \sim 10^{-1}$. Concerning the $\mu$ 
problem, it was shown in \cite{AltII} that the potential minimization conditions are similar to the ones in 
the NMSSM, with the substitution $N^c \leftrightarrow S$.

There exist another model \cite{AltI} aimed at solving both the neutrino mass and $\mu$ term problems. 
Within this model, three gauge-singlets are added to the MSSM superfield content: a right handed neutrino
$N_i^c$ giving rise to Dirac neutrino masses, a singlet $S$ addressing the $\mu$ naturalness ``\`a la NMSSM'',
and, a singlet $\Phi_i$ which is essential in order to drive simultaneously a spontaneous breaking of the 
R-parity and electroweak symmetries in a phenomenologically consistent way. In this framework, it is not clear
from the related literature \cite{AltI} what must be the typical neutrino Yukawa coupling values in order to
generate a physical neutrino mass scale around the $\mbox{eV}$.

We now compare our scenario, namely the NMSSM in the presence of the \rpv bilinear term $\mu_i L_i H_u$ ({\it c.f.} 
superpotential (\ref{WmI})), with the MSSM in the presence of this same bilinear term. The latter scenario, which 
suffers from the $\mu$ problem, was extensively studied in regard of the neutrino mass aspect \cite{PhysRep}.
\\ Let us consider a generic basis in which $v_i=\langle  \tilde \nu_i  \rangle  \neq  0$ and $\mu_i \neq 0$ simultaneously. 
Then, requiring a neutrino mass scale typically smaller than $1\mbox{eV}$ imposes the alignment \cite{AlignI,AlignII} 
of vectors $v_\alpha \equiv (v_d,v_i)$ and $\mu_\alpha \equiv (\mu,\mu_i)$ up to 
$$
\sin \zeta \lesssim 3 \ 10^{-6} \ \sqrt{1+\tan^2 \beta},
$$ 
where the basis-independent angle $\zeta$ is defined by 
\footnote{For a general discussion on basis-independent parametrization, see \cite{davidson}.},
$$
\cos \zeta = \frac{\sum_\alpha v_\alpha \mu_\alpha}{\sqrt{(\sum_\alpha v_\alpha^2)(\sum_\alpha \mu_\alpha^2)}}.
$$
Such an alignment arises naturally in the framework of horizontal symmetries, but it would then rely on the condition
$|\mu_i| \ll |\mu|$, or more precisely $|\mu_i / \mu| 
< {\cal O}(10^{-5})$ in the first explicit realization proposed in 
\cite{AlignII}. Once more, this hierarchy is more dramatic than in our scenario, where a ratio 
$|\mu_i/\mu| \simeq 10^{-1}$ allows a sufficient neutrino mass suppression relatively to the electroweak energy scale.
\\ Besides, in various accurate three-flavor analyzes \cite{MSSMrpvA}-\cite{MSSMrpvB}, 
it was shown that the combined tree and loop 
MSSM contributions can accommodate the experimental measurements on neutrino masses and leptonic mixing angles. 
In particular, complete scans of the parameter space \cite{AM1,AM2,abl} have shown that the basis-independent
quantities $\delta_\mu^i$ and $\delta_B^i$ (see \cite{DL1}) must be of order 
$|\delta_\mu^i| \sim 10^{-7}$ and $|\delta_B^i| \sim 10^{-5}$, 
assuming sparticle masses fixed at a common effective supersymmetry scale equal to $100\mbox{GeV}$. In the basis 
where $v_i=0$, these two quantities correspond respectively to the ratios $|\mu_i/\mu|$ and $|B_i/B|$
($\mu_i$ and $B_i$ can be negative), 
$B$ being the soft supersymmetry breaking parameter entering the scalar potential via the interaction 
$B h_u h_d$. So in this basis (that we have considered throughout the study of our scenario), the required ratio 
$|\mu_i/\mu| \sim 10^{-7}$ is much smaller than in our scenario where $|\mu_i/\mu|$ can reach $\sim 10^{-1}$,
with respect to the correct order of magnitude for the neutrino mass scale.
The trilinear \rpv terms, if included, do not change the order of magnitude of the ranges for $\delta_\mu^i$ 
and $\delta_B^i$, and, the \rpv trilinear coupling constants were found to be 
$\sim 10^{-4}$ to satisfy all constraints from neutrino data.

\section{Scenario II}
\label{versionII}

\subsection{Neutralino masses}

{\bf $\bullet$ Superpotential:}
We turn to a version of our scenario, proposed in Section \ref{versionI}, where the bilinear 
\rpv interactions have the same origin as the $\mu$ term: those are now generated through the
VEV of the scalar component of the $S$ singlet superfield. Indeed, let us assume that the bilinear 
\rpv interactions of Eq.(\ref{WmI}) are forbidden by a symmetry (exactly like a symmetry is 
imposed within the NMSSM in order to kill the term $\mu H_u H_d$). Then the following 
supersymmetric and gauge invariant term, which is renormalizable, generates $\mu_i$-like terms: 
\begin{equation}
W_{II} = W_{NMSSM} + \lambda_i S L_i H_u,
\label{WmII}
\end{equation}
where $\lambda_i$ are new dimensionless coupling constants. This trilinear 
term, which has no analog in the MSSM, could be rotated away, by an $SU(4)$ rotation on 
$(H_d,L_i)^T$, into the pure NMSSM term $S H_u H_d$. However, the trilinear term $S L_i H_u$
would be regenerated via the renormalization group equations (in the presence of L violating 
couplings) \cite{NMSSM,NMpheno,littleFT}. We also
note that no massless Goldstone boson (the problematic Majoron) appears when $s$ acquires a 
VEV, since the lepton number is already explicitly broken by the trilinear term of Eq.(\ref{WmII}).
This trilinear term also violates explicitly the R-parity symmetry, as the bilinear $\mu_i$ terms 
of superpotential (\ref{WmI}).
The existence of the other trilinear \rpv interactions depends on the superpotential symmetry.
In order to protect the proton against its possible decay channels, this symmetry could be a GLP 
(killing $\lambda_{i,j,k} L_i L_j E_k^c$, $\lambda'_{i,j,k} L_i Q_j D_k^c$ and $\lambda_i S L_i H_u$), 
a GBP (killing $\lambda''_{i,j,k} U_i^c D_j^c D_k^c$) or a GMP (forbidding both L and B violating 
trilinear terms). It is desirable that all the global symmetries of the superpotential are discrete 
gauge symmetries \cite{Ibanez}. Under this hypothesis, by imposing the non-trivial conditions of linear 
anomaly (except the gravitational one) cancellation on the original $Z_N$ cyclic local (R-)symmetries, 
the authors of \cite{Chemtob} have shown that some residual symmetries of the three types, GLP, GBP or 
GMP, are possible within the NMSSM.

{\bf $\bullet$ Mass matrix:}
In this new framework, the Lagrangian containing the neutralino masses is the identical as (\ref{LAGmass})
but with a different \rpv part of the mass matrix mixing neutrinos and neutralinos:
\bea
\xi^\prime_{\rpv} = 
\left( 
\begin{array}{ccccc}
0 & 0 & 0 & \lambda_1 \langle s \rangle  & \lambda_1 v_u  \\
0 & 0 & 0 & \lambda_2 \langle s \rangle  & \lambda_2 v_u   \\
0 & 0 & 0 & \lambda_3 \langle s \rangle  & \lambda_3 v_u   
\end{array}
\right).
\label{RPVmassBIS}
\eea
Note the presence of the new mixings between $\tilde s$ and $\nu_i$.

\subsection{Effective neutrino mass}

{\bf $\bullet$ Mass expression:}
Since we restrict to the situation $|\lambda_i/\lambda|<10^{-1}$, the effective neutrino mass matrix 
is still given in a good approximation by the see-saw formula:
\begin{equation}
m_{\nu} = - \xi^\prime_{\rpv} \ {\cal M}_{NMSSM}^{-1} \ \xi_{\rpv}^{\prime \ T}.
\label{seesawBIS}
\end{equation}
From Eq.(\ref{NMSSMmass}), Eq.(\ref{RPVmassBIS}) and Eq.(\ref{seesawBIS}), we derive analytically 
the following effective Majorana neutrino mass matrix,
\begin{equation}
m_{\nu i j} = \frac{\lambda_i}{\lambda} \frac{\lambda_j}{\lambda} 
\frac{M_1 M_2 (2 \kappa \mu / \lambda)}{Det({\cal M}_{NMSSM})}
\bigg (
\frac{\lambda^2 v_u v_d}{2 \kappa \mu / \lambda} + \frac{\mu}{2}
\bigg )
2 \mu
M_Z^2 [ \frac{\sin^2\theta_W}{M_1} + \frac{\cos^2\theta_W}{M_2} ] \cos^2 \beta.
\label{MnuEFFBIS}
\end{equation}

In this scenario, we remark in Eq.(\ref{MnuEFFBIS}) that the specific 
ratio particularly relevant for the discussion becomes $\lambda_i/\lambda$ instead of $\mu_i/\mu$
(as in scenario I), 
since one has here (in terms of effective $\mu_i$ and $\mu$ parameters): 
$$
\frac{\mu_i}{\mu}=\frac{\lambda_i \langle s \rangle }{\lambda \langle s \rangle }=\frac{\lambda_i}{\lambda}.
$$

{\bf $\bullet$ Origin of smallness:}
Once again, we see on the neutrino mass matrix (\ref{MnuEFFBIS}) that there is a possible
source of suppression from an approximate cancellation between the two terms in brackets. 
This neutrino mass suppression has a different source from the other suppression issued from 
the smallness of ratio $|\lambda_i/\lambda|$ (see Eq.(\ref{MnuEFFBIS})). The smallness of 
this ratio would provide an interpretation of the neutrino mass hierarchy problem by introducing 
another hierarchy, namely the hierarchy between the fundamental parameters $\lambda_i$ and 
$\lambda$.

\begin{figure}[t]
\begin{center}\unitlength1mm
\SetScale{2.8}
\begin{boldmath}
\begin{picture}(80,80)(0,-10)
\Line(0,0)(15,0)\Line(65,0)(15,0)
\Line(78,0)(65,0)
\DashLine(25,0)(25,30){2}
\Text(25,-4.5)[c]{$\lambda_i v_{u}$}
\DashLine(55,0)(55,30){2}
\Text(55,-4.5)[c]{$\lambda_j v_{u}$}
\Text(52.6,29.6)[l]{$\times$}
\Text(55,35)[c]{$\langle {h}^0_{u} \rangle $}
\Text(22.9,29.6)[l]{$\times$}
\Text(25,35)[c]{$\langle {h}^0_{u} \rangle $}
\Text(73,0)[r]{$<$}
\Text(-1,0)[r]{$\nu_i$}
\Text(8,0)[r]{$>$}
\Text(80,0)[l]{$\nu_j$}
\Text(40,-6)[c]{$\widetilde{s}$}
\Text(33,0)[r]{$<$}
\Text(50,0)[r]{$>$}
\Text(40,5)[c]{$2\kappa \mu/\lambda$}
\Text(40,0)[c]{$\times$}
\end{picture}
\end{boldmath}
\caption{Feynman diagram for the contribution to Majorana neutrino mass which arises in the 
NMSSM through the trilinear coupling of Eq.(\ref{WmII}).}
\protect\label{fig:diagramNMSSMBIS}
\end{center}
\end{figure}
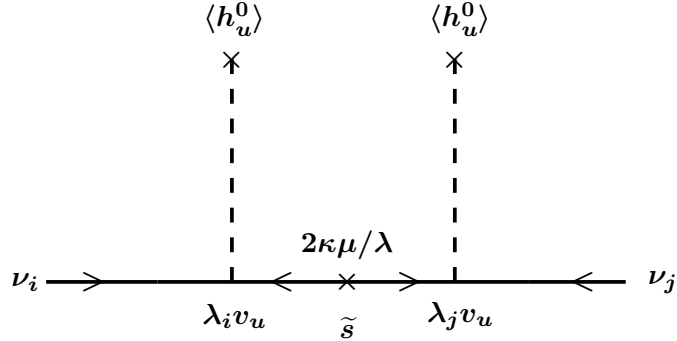

This possible cancellation in Eq.(\ref{MnuEFFBIS}) can be understood from a diagramatic point of view,
as before. Indeed, in the present framework, the Majorana neutrino mass still receives contributions 
from the previous exchanges of gauginos and singlino represented in diagrams (\ref{fig:diagramMSSM}), 
except that the $\mu_i$ mass insertions in these diagrams come 
now through the VEV of $s$ and should be parameterized instead by $\lambda_i \langle s \rangle$. 
Furthermore, there is 
new possible exchange of singlino which we have drawn in Fig.(\ref{fig:diagramNMSSMBIS}). This contribution
is due to the new trilinear \rpv term of superpotential (\ref{WmII}). The above approximate cancellation 
between the two terms in brackets entering neutrino mass (\ref{MnuEFFBIS}) would originate from a 
compensation between the two types of process contributing to neutrino mass: the exchange of gauginos 
(Fig.(\ref{fig:diagramMSSM})(a)) and the exchanges of a singlino (Fig.(\ref{fig:diagramMSSM})(b) and 
Fig.(\ref{fig:diagramNMSSMBIS})).

This cancellation-like source of neutrino mass suppression requires some fine tuning on NMSSM parameters. 
It is, once more, $\lambda$ that faces the strongest fine tuning. Nevertheless, this fine tuning on $\lambda$ 
decreases greatly as $| \mu |$ and $| \kappa |$ ($\lambda$ and $\tan \beta$) get smaller (larger).

\subsection{Numerical results and discussion}

\begin{table}[t]
\begin{center}
\begin{tabular}{c|c|c|c|c|c|c|c}
 & $\lambda$ & $\mu$ & $\tan \beta$ & $M_1$ & $M_2$ & $\lambda_1$ & $- \kappa$  
\\
&  & $\mbox{[GeV]}$ &  & $\mbox{[GeV]}$ & $\mbox{[GeV]}$ & &
\\
\hline
A & 0.7 &  110 &  50   & 100 & 100 & $7 \ 10^{-2}$ & $(1.71783 - 1.7177) \ 10^{-2}$ 
\\
&  &   &    &  &  & $7 \ 10^{-4}$ & $(16.2 - 7) \ 10^{-3}$ 
\\
\hline
B & 0.7 &  300 &  50   & 50 & 500 & $7 \ 10^{-2}$ & $(2.30954 - 2.3095) \ 10^{-3}$ 
\\
&  &   &   & & & $7 \ 10^{-4}$ & $(2.26 - 1.8) \ 10^{-3}$ 
\\
\hline
C & 0.4 &  -110 &  30   & 100 & 100 & $4 \ 10^{-2}$ & $(5.33839 - 5.3385) \ 10^{-3}$ 
\\
& &  &   &  &  & $4 \ 10^{-4}$ & $(5.41 - 6) \ 10^{-3}$ 
\end{tabular}
\caption{Sets (A,B,C) of values, for the parameters entering the whole neutralino mass matrix, 
which reproduce the correct neutrino mass. The two values of parameter $\kappa$ correspond to 
the neutrino masses $m_{\nu}= 0.1 \mbox{eV} - 1 \mbox{eV}$ ($m_{\nu}$ being defined 
via Eq.(\ref{MnuEFFBIS})), respectively. As the one-flavor case is considered here, 
the flavor index $i$ of \rpv coupling constant $\lambda_i$ takes only the value $i=1$.}
\label{tab:parII}
\end{center}
\end{table}

{\bf $\bullet$ Neutrino mass suppression:}
In Table \ref{tab:parII}, we show characteristic sets of parameters for which the neutrino mass 
(\ref{MnuEFFBIS}) at one flavor is equal to $0.1 \mbox{eV}$ and $1 \mbox{eV}$, covering the 
range of values motivated by experimental data (Eq.(\ref{bdTOT})).

In Table \ref{tab:parII}, the $\kappa$ value is fixed by the other parameters via formula (\ref{MnuEFFBIS}). 
Here, we have chosen to fix $\kappa$ as it is direct to solve Eq.((\ref{MnuEFFBIS})) in term of this parameter. 

Let us comment on the results in Table \ref{tab:parII}. $\kappa$ is chosen negative so that 
the cancellation between the terms in brackets of expression (\ref{MnuEFFBIS}) is effective. 
This cancellation can be only partially responsible for the neutrino mass reduction, as for 
points in the table with $| \lambda_1/\lambda | = 10^{-3}$ ($\lambda_1 = 4,7 \ 10^{-4}$). 
This cancellation can also be the principal mechanism that suppresses the neutrino mass,
as for the points with $| \lambda_1/\lambda | = 10^{-1}$ ($\lambda_1 = 4,7 \ 10^{-2}$).
Then the wanted neutrino mass suppression is reached without requiring any highly 
hierarchical pattern.

\begin{table}[t]
\begin{center}
\begin{tabular}{c|c|c|c}
& $m_{\tilde \chi^0_1}$  & $m_{\tilde \chi^0_5}$  &  ${\cal F}_\lambda$  
\\
& $\mbox{[GeV]}$ & $\mbox{[GeV]}$ &
\\
\hline
A & 34 & 193 &   $(0.2 - 1.9) \ 10^{-5}$ 
\\
& 33 & 193 & $(2.1 - 196) \ 10^{-2}$   
\\
\hline
B & 7 & 520 &  $(0.7 - 6.9) \ 10^{-6}$ 
\\
& 7 & 520  & $(0.7 - 9.4) \ 10^{-2}$   
\\
\hline
C & 18 & 180  &  $(0.4 - 4.1) \ 10^{-6}$ 
\\
& 18 & 180 & $(0.4 - 3.3) \ 10^{-2}$ 
\end{tabular}
\caption{Lowest [$m_{\tilde \chi_1^0}$] and highest [$m_{\tilde \chi_5^0}$] neutralino masses  
for the points A,B,C of parameter space presented in Table \ref{tab:parII}. Together with these 
masses, we also give the value of fine tuning quantity ${\cal F}_\lambda$ defined in text for the 
$\lambda$ parameter. The two values of ${\cal F}_\lambda$ correspond respectively to the two 
$\kappa$ values in Table \ref{tab:parII} (leading to $m_{\nu}= 0.1 \mbox{eV} - 1 \mbox{eV}$). 
For each point, the first and second lines are respectively associated to $\lambda_1=4,7 \ 10^{-2}$ 
and $\lambda_1=4,7 \ 10^{-4}$, as in Table \ref{tab:parII}.}
\label{tab:FTII}
\end{center}
\end{table}

{\bf $\bullet$ Fine tuning:}
We quantify the fine tuning on $\lambda$ with variable (\ref{FTratio}), now defined with
the neutrino mass (\ref{MnuEFFBIS}). On Table \ref{tab:FTII}, we give the values of this variable 
${\cal F}_\lambda$ for the points of parameter space shown in Table \ref{tab:parII} which reproduce 
the correct neutrino masses through the compensation mechanism. A comparison of points A and B in 
Table \ref{tab:FTII} shows that the fine tuning on $\lambda$ is softer if $| \mu |$ decreases. 
In the same way, by comparing parameters A and C, one observes that the fine tuning is significantly 
improved for larger values of $\lambda$ or $\tan \beta$. Table \ref{tab:FTII} also exhibits that the 
fine tuning is soften for higher neutrino masses and smaller $| \lambda_1/\lambda |$ ratios.

{\bf $\bullet$ Tachyons:}
Unfortunately, it turns out that for any domain of the parameter space $\{ \tan \beta, \mu, M_1, M_2 \}$, 
the fact of requiring the neutrino mass (\ref{MnuEFFBIS}) to be suppressed down to the $\mbox{eV}$ scale,
at least partially through our cancellation mechanism (namely for $| \lambda_i/\lambda | \gtrsim 10^{-6}$), 
imposes the ratio $| \kappa / \lambda |$ to be small, leading to the occurrence of unacceptable tachyons 
in the CP-even Higgs sector.

We have checked this feature of our scenario by using the code NMHDECAY \cite{nmhdecay}
which applies on the pure NMSSM 
parameter space. However, this procedure is believed to be consistent since the additional trilinear 
\rpv interaction $\lambda_i S L_i H_u$ in the superpotential (see Eq.(\ref{WmII})) is not expected to 
induce considerable modifications in the scalar potential of the NMSSM. As a matter of fact, we
have systematically restricted ourselves to the case $| \lambda_i/\lambda | \leq 10^{-1}$ 
({\it c.f.} Eq.(\ref{WNMSSM})).

In a situation where the $| \lambda_i/\lambda |$ ratio would be of order unity or even larger, 
giving rise to important changes in the NMSSM potential, it could happen that our cancellation 
mechanism for neutrino mass suppression would be active without implying necessarily the 
appearance of tachyons in the theory.

Another way out of this theoretical problem is to focus on the particular case $A_\kappa = 0$,
in which no tachyons emerge from the CP-even sector. This possibility is conceivable as the 
trilinear soft supersymmetry breaking parameter $A_\kappa$, which was previously introduced in 
Section \ref{NumResI}, does not affect the neutralino mass matrix (\ref{CHImass}) on which is 
based our analysis.

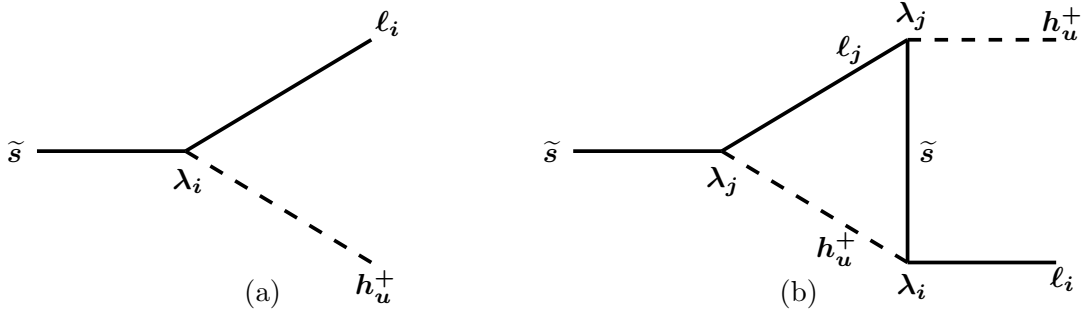
\begin{figure}[t]
\unitlength 1mm
\SetScale{2.8}
\begin{boldmath}
\begin{center}
\begin{picture}(60,30)(0,-10)
\Line(0,0)(20,0)
\Line(45,15)(20,0)
\DashLine(45,-15)(20,0){2}
\Text(-2,0)[r]{$\widetilde s$}
\Text(22,-4)[r]{$\lambda_i$}
\Text(45,17)[l]{$\ell_i$}
\Text(45,-18)[c]{$h^+_{u}$}
\Text(30,-19)[c]{(a)}
\end{picture}
\hspace{1cm}\begin{picture}(60,30)(0,-10)
\Line(0,0)(20,0)
\Line(45,15)(20,0)
\DashLine(45,-15)(20,0){2}
\Line(45,15)(45,-15)
\DashLine(45,15)(65,15){2}
\Line(45,-15)(65,-15)
\Text(-2,0)[r]{$\widetilde s$}
\Text(22,-4)[r]{$\lambda_j$}
\Text(35,13.5)[l]{$\ell_j$}
\Text(35,-13)[c]{$h^+_{u}$}
\Text(45,-18)[c]{$\lambda_i$}
\Text(45,18)[c]{$\lambda_j$}
\Text(30,-19)[c]{(b)}
\Text(47,0)[c]{$\widetilde s$}
\Text(65,17)[c]{$h^+_{u}$}
\Text(65,-17)[c]{$\ell_i$}
\end{picture}
\end{center}
\end{boldmath}\vskip 0.3cm
\caption{CP-asymmetry decay diagrams at tree (a) and loop (b) level in the NMSSM with
\rpv trilinear couplings.}
\protect\label{lepto}
\end{figure}

{\bf $\bullet$ Leptogenesis:}
Anyway, within this second scenario, the suppression of neutrino mass scale can 
be insured by a small $| \lambda_i/\lambda |$ value, so that the tachyonic regions associated to the 
cancellation mechanism are avoided. Then this scenario still possesses a new interesting phenomenological 
feature: the extra singlet of the NMSSM can produce, via decay channels involving \rpv trilinear couplings, 
a thermal leptogenesis. This leptogenesis can be converted into the baryonic sector through sphaleron induced 
processes, explaining then the baryon asymmetry of the universe. The lepton asymmetry arises through the 
out-of-equilibrium decay of the singlet in a L and CP-violating way, according to Sakharov's constraints  
\cite{Sakharov}.
Indeed, the CP-asymmetry may be generated from the interference between the tree level diagram of Fig.(\ref{lepto})(a)
and one-loop diagrams such as the one drawn in Fig.(\ref{lepto})(b). Only the \rpv trilinear couplings
of Eq.(\ref{WmII}) enter the two diagrams in Fig.(\ref{lepto}). In that case, the CP-asymmetry
\bea
\epsilon={\Gamma(S\to \ell H)-\Gamma(S\to \bar \ell H)\over \Gamma(S\to \ell H)+\Gamma(S\to \bar \ell  H)}
\eea
would be proportional to $\sum_j \lambda_i^2 \lambda_j^2 f$, 
where $f$ is the loop 
function. There exist other types of diagrams, generating a CP-asymmetry, which involve the lepton Yukawa
($Y^\ell_{ij}$) and \rpv trilinear ($\lambda_i$) coupling constants.

\section{Conclusion}
\label{conclu}

First, we have considered the NMSSM, which solves the $\mu$ problem, in the presence of bilinear 
\rpv interactions $\mu_i L_i H_u$ (scenario I). 
In this context, we have found that a cancellation mechanism 
arises for suppressing the Majorana neutrino mass and thus provides an interpretation to the smallness
of neutrino mass compared to the electroweak scale. This mechanism, which relies on the existence of 
the gauge-singlet $S$ introduced by the NMSSM, offers a solution for the neutrino mass problem 
which is interestingly connected to the solution for the $\mu$ problem.
\\
More precisely, by using the NMHDECAY program, we have obtained various characteristic points of 
the NMSSM parameter space which satisfy the experimental constraints from collider physics, fulfill the 
theoretical consistency conditions (physical minimum, no Landau singularity,\dots) and simultaneously
{\it generate} neutrino masses of order of the $\mbox{eV}$ scale through our cancellation mechanism. 
By the verb `generate', we mean here that small neutrino mass values are effectively produced 
without introducing a strong hierarchy between the fundamental parameters. Indeed, in the basis where
$\langle  \tilde \nu_i  \rangle =0$, the obtained parameters lead to neutrino masses $m_\nu \in [0.1,1] \mbox{eV}$ 
with $10^{-3} \lesssim |\mu_i / \mu| \lesssim 10^{-1}$ (the extreme values given here, for the ranges of 
neutrino mass and $|\mu_i / \mu|$ ratio, are not corresponding to each other). 
\\
In comparison, the see-saw mechanism suppresses sufficiently neutrino masses by 
introducing an high hierarchy between the Dirac and Majorana masses. Furthermore, in the
MSSM with a non-vanishing $\mu_i L_i H_u$ term, realistic neutrino masses are achieved for
$|\mu_i / \mu| \sim 10^{-7}$ typically. Finally, in the new version of the MSSM suggested recently 
in \cite{AltII}, which constitutes an alternative to our scenario as it addresses both the $\mu$ 
value and neutrino mass problems, a stronger hierarchy of $\sim 10^{-6}$ is required between the 
the neutrino and top quark Yukawa coupling constants. 
\\
Nevertheless, our new cancellation mechanism for neutrino mass suppression needs a certain fine tuning 
on some NMSSM parameters. For some of the obtained parameters mentioned above, that generate neutrino
masses around the $\mbox{eV}$, the most important fine tuning reaches the acceptable level of 
$\sim 3 \ 10^{-1}$ ($\sim 10^{-3}$) for $|\mu_i / \mu| \simeq 10^{-3}$ ($\simeq 10^{-1}$).
\\ 
The continuation of this study \cite{prep} would be the combination 
of tree and one-loop contributions with three flavors 
in order to accommodate all the last data on neutrino masses and leptonic mixing angles.

Secondly, we have studied another attractive version of this model (scenario II), namely the NMSSM with 
\rpv $\mu_i$-like interactions generated naturally by the VEV of the $S$ scalar component, through 
the trilinear term $\lambda_i S L_i H_u$. There the same kind of cancellation mechanism can occur for
the neutrino mass suppression. Based on this mechanism, we have easily found parameters which give rise 
to $m_\nu \in [0.1,1] \mbox{eV}$ for $10^{-3} \leq |\lambda_i / \lambda| \leq 10^{-1}$ corresponding
to a quite soft hierarchy. The associated fine tuning can reach $\sim 1$ ($\sim 10^{-5}$) for 
$|\lambda_i / \lambda| = 10^{-3}$ ($= 10^{-1}$). However, here, the cancellation mechanism seems to 
imply the occurrence of tachyons in the CP-even sector, at least in the simplest form of the NMSSM.
So one should think of some way out, like restricting to the particular situation where $A_\kappa$ 
vanishes.

It would also be interesting to find an independent theoretical reason for the compensation, between 
the two types of process exchanging the gauginos and singlino, which explains this new cancellation 
mechanism of neutrino mass suppression. In the same philosophy as for the see-saw mechanism, where
the Dirac/Majorana mass hierarchy introduced finds a natural realization within the framework of the 
$SO(10)$ GUT model.

Let us finish by commenting on the specific and rich phenomenology of the NMSSM with additional 
\rpv $\mu_i$-like interactions. We have discussed the fact that such a framework opens the possibility 
of new leptogenesis scenarios. This framework also leads to new decay channels for the Lightest 
Supersymmetric Particle (LSP). For instance, in the case where the LSP is the lightest neutralino,
it can decays as $\tilde \chi^0_1 \to \nu_i Z^0$ and $\tilde \chi^0_1 \to l_j^\pm W^\mp$ via the 
\rpv $\mu_i$-like mixings $\tilde h^0_u \nu_i$ or $\tilde s \nu_i$. The value of the LSP life time associated to 
these new decays
\footnote{In our scenario, the $|\mu_i/\mu|$ ratio can reach values close to unity which tends to
increase significantly the width of these new LSP decays, except if the $\tilde \chi^0_1$ is mainly 
composed by $\tilde B^0$, $\tilde W^0_3$ and/or $\tilde h^0_d$.}
is fundamental in regard of the collider physics (if the LSP decays inside the
detectors, the typical supersymmetric signatures are multi-jets/leptons instead of missing energy) 
as well as of the dark matter problem (the LSP remains a good WIMP candidate only if it is stable,  
relatively to the age of the universe).

\section*{Acknowledgments}

The authors are grateful to A.~Djouadi for stimulating discussions and 
G.~Bhattacharyya for reading 
the manuscript. It is also a pleasure to thank M.~Chemtob, U.~Ellwanger, P.~Fayet, 
C.~Hugonie, C.~Mun\~oz and C.~A.~Savoy for fruitful conversations.

\end{document}